\documentclass[amsmath,amssymb,showpacs,preprint]{revtex4}

\usepackage{epsfig}
\usepackage{graphicx}

\begin{document}

\title{Anisotropic inflation in Finsler spacetime}

\author{Xin Li $^{1,2}$}
\email{lixin1981@cqu.edu.cn}
\author{Sai Wang $^2$}
\email{wangsai@itp.ac.cn}
\author{Zhe Chang $^{2,3}$}
\email{changz@ihep.ac.cn}
\affiliation{$^1$Department of Physics, Chongqing University, Chongqing 401331, China\\
$^2$State Key Laboratory Theoretical Physics, Institute of Theoretical Physics, Chinese
Academy of Sciences, Beijing 100190, China\\
$^3$Institute of High Energy Physics, Chinese Academy of Sciences, Beijing 100049, China}

\begin{abstract}
We suggest the universe is Finslerian in the stage of inflation. The Finslerian background spacetime breaks rotational symmetry and induces parity violation. The primordial power spectrum is given for quantum fluctuation of the inflation field. It depends not only on the magnitude of wavenumber but also on the preferred direction. We derive the gravitational field equations in the perturbed Finslerian background spacetime, and obtain a conserved quantity outside the Hubble horizon. The angular correlation coefficients are presented in our anisotropic inflation model. The parity violation feature of Finslerian background spacetime requires that the anisotropic effect only appears in angular correlation coefficients if $l'=l+1$. The numerical results of the angular correlation coefficients are given to describe the anisotropic effect.
\end{abstract}

\maketitle
\section{Introduction}
Cosmic inflation \cite{Starobinsky}, as one of basic ideas of modern cosmology, plays an essential role for the very early universe. During inflation era, the universe undergo an exponential expansion in a very short time. Such a property of inflation makes it successfully accounting for the flatness problem and the horizon problem \cite{Riotto}. Moreover, the vacuum fluctuation of the inflation field that generates the primordial energy density perturbation is a seed of cosmic anisotropy. Associating with the primordial energy density perturbation, the initial scalar perturbation is predicted to be adiabatic, Gaussian, and nearly scale-invariant. The recent astronomical observations on the anisotropy of cosmic microwave background (CMB) \cite{CMB1,CMB2} show that the exact scale invariance of the scalar perturbation is broken with more than 5 standard deviations. And the observations give stringent limits on deviation from Gaussian statistics \cite{No gauss}. These facts support that the inflation model is a preferred model that generates the primordial quantum fluctuation.
In the standard single-field inflation model, cosmic inflation can be described by nearly de Sitter (dS) spacetime. Only six isometries are preserved in nearly dS spacetime, namely, three spatial translations and three spatial rotations. Other four isometries of dS spacetime are broken. Especially, the violation of time translation yields that the primordial power spectrum is not scale-invariant. The observations on CMB anisotropy \cite{CMB1,CMB2} limit the magnitude of deviation from the scale invariant to be $\mathcal{O}(10^{-2})$.

Recently, the CMB power asymmetry has been reported \cite{Power asymmetry,Akrami:2014eta}. It encourages physicists to study the anisotropic inflation model which breaks the rotational symmetry of the nearly dS spacetime. In the usual anisotropic inflation model \cite{Vector model}, a primordial vector field aligned in a preferred direction is involved and the rotational symmetry is broken. In such anisotropic inflation model, the comoving curvature perturbation becomes statistically anisotropic \cite{Vector model1}. Usually, the background spacetime of anisotropic inflation model is described by Bianchi spacetime \cite{Bianchi spacetime}. The Planck 2013 CMB temperature map gives a limit on the deviation of Bianchi spacetime from Friedmann-Robertson-Walker (FRW) spacetime. The upper bound of rotation invariance violation during inflation is limited to be less than $\mathcal{O}(10^{-9})$ \cite{Komatsu limit}.

In this paper, we will study anisotropic inflation in which the background spacetime is taken to be Finslerian. Finsler geometry \cite{Book by Bao} is a new geometry which involves Riemann geometry as its special case. Chern pointed out that Finsler geometry is just Riemann geometry without quadratic restriction, in his Notices of AMS. We choose Finsler geometry to investigate anisotropic inflation for three reasons. First, the flat Finsler spacetime, as the counterpart of Minkowski spacetime, admits less Killing vectors than Riemann spacetime does \cite{Finsler PF}. The numbers of independent Killing vectors of an $n$ dimensional non-Riemannian Finsler spacetime should be no more than $n(n-1)/2+1$ \cite{Wang}. It is expected that the rotational symmetry is broken in Finsler spacetime. Second, the gravitational field equation in the modified Finslerian FRW spacetime has been setup \cite{Finslerian dipole}. Such a Finsler spacetime can be used to investigate the cosmological preferred direction that is implied by the Union2 SnIa data \cite{Perivolaropoulos}. Last but not the least, generally, Finsler spacetime is non-reversible under parity flip, $x\rightarrow-x$. A typical non-reversible Finsler spacetime is Randers spacetime \cite{Randers}. Such a property makes a function in Fourier space $\phi(-\vec{k})$ to be different from $\phi^\ast(\vec{k})$. Thus, one can expect that the primordial power spectrum in Finsler spacetime violates the parity symmetry, namely, the power spectrum is not invariant under the parity flip, $k\rightarrow-k$.

The rest of the paper is arranged as follows: In Section \ref{sec:anisotropy}, we present the anisotropic inflation model in Finsler spacetime. The background Finsler spacetime breaks the rotational symmetry and induces parity violation. We derive the primordial power spectrum for the quantum fluctuation of inflation field. In Section \ref{sec:field eq}, we derive the gravitational field equations in the perturbed Finslerian background spacetime, and obtain a conserved quantity outside the Hubble horizon. In section \ref{sec:spectrum}, we derive the angular correlation coefficients in our anisotropic inflation model. And we plot the numerical results of the angular correlation coefficients that describes the anisotropic effect. Conclusions and remarks are given in Section \ref{sec:conclusion}.

\section{Anisotropic inflation in Finsler spacetime}\label{sec:anisotropy}
Finsler geometry is based on the so called Finsler structure $F$ defined on the tangent bundle of a manifold $M$, with the property $F(x,\lambda y)=\lambda F(x,y)$ for all $\lambda>0$, where $x\in M$ represents position and $y$ represents velocity. The Finslerian metric is given as \cite{Book by Bao}
\begin{equation}
g_{\mu\nu}\equiv\frac{\partial}{\partial
y^\mu}\frac{\partial}{\partial y^\nu}\left(\frac{1}{2}F^2\right).
\end{equation}
Throughout this paper, the indices are lowered and raised by $g_{\mu\nu}$ and its inverse matrix $g^{\mu\nu}$.
The Finslerian metric reduces to Riemannian metric, if $F^2$ is quadratic in $y$.

In order to describe the anisotropic inflation, we propose that the background Finsler spacetime is of the form
\begin{equation}\label{FRW like}
F^2_0=y^ty^t-a^2(t)F_{Ra}^2,
\end{equation}
where $F_{Ra}$ is a Randers space \cite{Randers}
\begin{equation}
F_{Ra}=\alpha+\beta=\sqrt{\delta_{ij}y^iy^j}+\delta_{ij}b^iy^j,
\end{equation}
$\alpha$ denotes the Riemannian metric and $\beta$ denotes the one form.
Here, we require that the vector $b^i$ in $F_{Ra}$ is of the form $b^i=\{0,0,b\}$ and $b$ is a constant. It is obvious that the background spacetime (\ref{FRW like}) returns to FRW spacetime when $b=0$. The spatial part of the universe $F_{Ra}$ is a flat Finsler space, since all types of Finslerian curvature vanishes for $F_{Ra}$. The Killing equations of Randers space $F_{Ra}$ are given as \cite{Finsler PF,Finsler BH}
\begin{eqnarray}\label{Killing eq1}
L_V \alpha &=&\delta_{kj}\frac{\partial V^k}{\partial x^i}+\delta_{ki}\frac{\partial V^k}{\partial x^j}= 0,\\ \label{Killing eq2}
L_V \beta &=&b\frac{\partial V^z}{\partial x^i}= 0,
\end{eqnarray}
where $L_V$ is the Lie derivative along the Killing vector $V$. Noticing that the vector $b^i$ parallels with $z$-axis, we find from the Killing equations (\ref{Killing eq1},\ref{Killing eq2}) that there are four independent Killing vectors in Randers space $F_{Ra}$. Three of them represent the translation symmetry, and the rest one represents the rotational symmetry in $x-y$ plane. It means that the rotational symmetry in $x-z$ and $y-z$ plane are broken. Also, the spatial part of Finsler spacetime (\ref{FRW like}) is non-reversible for $z\rightarrow-z$. It means that the party is violated in Finsler spacetime (\ref{FRW like}).

In our anisotropic inflation model, we just consider the single-field ``slow-roll'' inflation. The action for this scalar field is given by
\begin{equation}\label{action}
S=\int d^4x\sqrt{-g}\left(\frac{1}{2}g^{\mu\nu}\partial_\mu\phi\partial_\nu\phi-V(\phi)\right).
\end{equation}
The quantum fluctuation \cite{Rev inflaton} is associated with the first order perturbation $\delta\phi(t,\vec{x})$ of the inflation field $\phi$. Noticing that the determinant of the background spacetime (\ref{FRW like}) is given by $g=-a^3F_{Ra}^3/(\delta_{ij}y^iy^j)^{3/2}$, we find from the action (\ref{action}) that the equation of motion for $\delta\phi$ is given as
\begin{equation}\label{EOM}
\delta\ddot{\phi}+3H\delta\dot{\phi}-a^{-2}\bar{g}^{ij}\partial_i\partial_j\delta\phi=0,
\end{equation}
where the dot denotes the derivative with respect to time, $H\equiv\dot{a}/a$ and $\bar{g}^{ij}$ is Finslerian metric for Rander space $F_{Ra}$. The Finslerian metric $\bar{g}^{ij}$ is a homogenous function of degree $0$ with respect to variable $y$. It means that the Finslerian metric $\bar{g}^{ij}$ depends on direction. Such a property makes the solution of the equation of motion (\ref{EOM}) much more complicated. In this paper, for simplicity, we take the direction of $y$ to be parallel with the wave vector $\vec{k}$ in the momentum space. Therefore, in the momentum space, the equation (\ref{EOM}) can be simplified as
\begin{equation}\label{EOM1}
\delta\ddot{\phi}+3H\delta\dot{\phi}-a^{-2}k^2_e\delta\phi=0,
\end{equation}
where the effective wavenumber $k_e$ is given by
\begin{equation}
k_e^2=\bar{g}^{ij}k_ik_j=k^2(1+b\hat{k}\cdot\hat{n}_z)^2.
\end{equation}
Then, following the standard quantization process in the inflation model \cite{Mukhanov book}, we can obtain the primordial power spectrum from the solution of the equation (\ref{EOM1}). It is of the form
\begin{equation}\label{spectrum delta phi}
\mathcal{P}_{\delta\phi}(\vec{k})=\mathcal{P}_0(k)\left(\frac{k}{k_e}\right)^3\simeq\mathcal{P}_0(1-3b\hat{k}\cdot\hat{n}_z),
\end{equation}
where $\mathcal{P}_0$ is an isotropic power spectrum for $\delta\phi$ which depends only on the magnitude of wavenumber $k$.
The term $3b\hat{k}\cdot\hat{n}_z$ in the primordial power spectrum $\mathcal{P}_{\delta\phi}$ represents the effect of rotational symmetry breaking.

\section{Gravitational field equation in anisotropic inflation}\label{sec:field eq}
At the end stage of inflation, the background metric perturbation is coupled with the inflation field. Thus the initial power spectrum differs from $\mathcal{P}_{\delta\phi}$.
In standard inflation model, the comoving curvature perturbation that links inflation field with background metric perturbation is conserved outside the Hubble horizon. In our anisotropic inflation model, we should find a counterpart of comoving curvature perturbation.

We start from investigating the gravitational field equation for the perturbation of Finsler structure (\ref{FRW like}). In this paper, we just consider scalar perturbation. For simplicity, the scalar perturbed Finsler structure is of the form
\begin{equation}\label{FRW like perturbed}
F^2=(1+2\Psi(t,\vec{x}))y^ty^t-a^2(t)(1+2\Phi(t,\vec{x}))F_{Ra}^2,
\end{equation}
where $\Psi$ and $\Phi$ are scalar perturbations.
In Finsler geometry, there is geometrical invariant quantity, i.e., Ricci scalar. It is of the form \cite{Book by Bao}
\begin{equation}\label{Ricci scalar}
Ric\equiv\frac{1}{F^2}\left(2\frac{\partial G^\mu}{\partial x^\mu}-y^\lambda\frac{\partial^2 G^\mu}{\partial x^\lambda\partial y^\mu}+2G^\lambda\frac{\partial^2 G^\mu}{\partial y^\lambda\partial y^\mu}-\frac{\partial G^\mu}{\partial y^\lambda}\frac{\partial G^\lambda}{\partial y^\mu}\right),
\end{equation}
where $G^\mu$ is geodesic spray coefficients
\begin{equation}
\label{geodesic spray}
G^\mu=\frac{1}{4}g^{\mu\nu}\left(\frac{\partial^2 F^2}{\partial x^\lambda \partial y^\nu}y^\lambda-\frac{\partial F^2}{\partial x^\nu}\right).
\end{equation}
The Ricci scalar only depends on the Finsler structure $F$ and is insensitive to connections. Plugging the perturbed Finsler structure (\ref{FRW like perturbed}) into the formula of Ricci scalar (\ref{Ricci scalar}), we obtain that
\begin{eqnarray}
F^2Ric&=&-3\frac{\ddot{a}}{a}y^ty^t+(a\ddot{a}+2\dot{a}^2)F^2_{Ra}+y^ty^t\left(a^{-2}\bar{g}^{ij}\Psi_{,i,j}-3\ddot{\Phi}+3H(\dot{\Psi}-2\dot{\Phi})\right)\nonumber\\
&&+y^ty^i\left(4H\Psi_{,i}-4\dot{\Phi}_{,i}\right)-y^iy^j\left(\Psi_{,i,j}+\Phi_{,i,j}\right)\nonumber\\
&&+F_{Ra}^2\left((2\dot{a}^2+a\ddot{a})(2\Phi-2\Psi)+a\dot{a}(6\dot{\Phi}-\dot{\Psi})+a^2\ddot{\Phi}-\bar{g}^{ij}\Phi_{,i,j}\right)\nonumber\\
&&+\frac{1}{2}\frac{\partial\bar{g}^{ij}}{\partial y^i}\left(\Phi_{,j,k}y^k+\dot{\Phi}_{,j}y^t+2H\Psi_{,j}y^t\right)F_{Ra}^2\nonumber\\
\label{Ricci scalar1}
&&-\frac{1}{2a^2}\frac{\partial\bar{g}^{ij}}{\partial y^i}\left(\Psi_{,j,k}y^k+\dot{\Psi}_{,j}y^t+2H\Psi_{,j}y^t\right)y^ty^t,
\end{eqnarray}
where the comma denotes the derivative with respect to spatial coordinate $x$ and $H\equiv\dot{a}/a$. In section \ref{sec:anisotropy}, we have taken the direction of $y$ to be parallel with the wave vector $\vec{k}$ in momentum space. And noticing the relation $\frac{\partial\bar{g}^{ij}}{\partial y^i}y_j=0$, we find that the term of equation (\ref{Ricci scalar1}) which is proportional to $\frac{\partial\bar{g}^{ij}}{\partial y^i}$ should vanish in momentum space. In reference \cite{Finslerian dipole,Finsler BH}, we have proved that the gravitational field equation in Finsler spacetime is of the form
\begin{equation}\label{field equation}
G^\mu_\nu=8\pi G T^\mu_\nu,
\end{equation}
where the modified Einstein tensor in Finsler spacetime is defined as
\begin{equation}\label{Einstein tensor}
G^\mu_\nu\equiv Ric^\mu_\nu-\frac{1}{2}\delta^\mu_\nu S,
\end{equation}
and $T^\mu_\nu$ is the energy-momentum tensor.
Here the Ricci tensor is defined as \cite{Akbar}
\begin{equation}\label{Ricci tensor}
Ric_{\mu\nu}=\frac{\partial^2\left(\frac{1}{2}F^2 Ric\right)}{\partial y^\mu\partial y^\nu},
\end{equation}
and the scalar curvature in Finsler spacetime is given as $S=g^{\mu\nu}Ric_{\mu\nu}$.
Plugging the equation for Ricci scalar (\ref{Ricci scalar1}) into the gravitational field equation, we obtain the background equations
\begin{eqnarray}\label{field eq t}
3H^2&=&8\pi GT^0_{(0)0},\\
\label{field eq i}
\frac{2\ddot{a}}{a}+H^2&=&-8\pi G T^i_{(0)i}/3,
\end{eqnarray}
and the perturbed equations in momentum space
\begin{eqnarray}\label{perturbe eq tt}
6H\dot{\Phi}-6H^2\Psi+2\frac{k_e^2\Phi}{a^2}&=&8\pi G\delta T^0_0,\\
\label{perturbe eq ti}
-i\left(2H\Psi-2\dot{\Phi}\right)k_i&=&8\pi G\delta T^0_i,\\
\label{perturbe eq ii}
2\ddot{\Phi}+H\left(6\dot{\Phi}-2\dot{\Psi}\right)-\Psi\left(2H^2+4\frac{\ddot{a}}{a}\right)+\frac{k_e^2}{a^2}\left(\Phi+\Psi\right)&=&8\pi G\delta T^i_i/3,\\
\label{perturbe eq ij}
-a^{-2}k^ik_j\left(\Phi+\Psi\right)&=&8\pi G\delta T^i_j~~~(i\neq j),
\end{eqnarray}
where $k^i=\bar{g}^{ij}k_j$. Here, the energy-momentum tensor in the above field equations is derived by the variation of the action (\ref{action}) of inflation field $\phi$. It is of the form
\begin{equation}
T_{\mu\nu}=\partial_\mu\phi\partial_\nu\phi-g_{\mu\nu}\left(\frac{1}{2}g^{\alpha\beta}\partial_\alpha\phi\partial_\beta\phi-V(\phi)\right),
\end{equation}
where $T^0_{(0)0}$ and $T^i_{(0)i}$ in equations (\ref{field eq t},\ref{field eq i}) correspond to zero-order part of inflation field, i.e. $\phi_0(t)$, and $\delta T^\mu_\nu$ in equations (\ref{perturbe eq tt},\ref{perturbe eq ti},\ref{perturbe eq ii},\ref{perturbe eq ij}) corresponds to first-order part of inflation field, i.e. $\delta\phi(t,\vec{x})$. Now, we find that the background equations (\ref{field eq t},\ref{field eq i}) are the same with the standard inflation model that means our universe is exponentially expanding if the slow-roll condition $\dot{\phi}_0\ll V(\phi_0)$ is satisfied. However, one should notice that there is difference between standard inflation model and our anisotropic inflation model. During the expansion, the spatial form of universe is Euclidean in standard inflation model. And in our anisotropic inflation model, the spatial form of universe is Finslerian. This anisotropic feature is obvious in the perturbed equations. The perturbed equations (\ref{perturbe eq tt},\ref{perturbe eq ti},\ref{perturbe eq ii},\ref{perturbe eq ij}) are the same with the standard inflation model except for replacing wavenumber $k$ with effective wavenumber $k_e$. $k_e$ depends not only on the magnitude of $k$ but also the preferred direction $\hat{n}_z$ that induces rotational symmetry breaking.

Following the approach of standard inflation model \cite{Mukhanov book}, we find from the field equations (\ref{field eq t}-\ref{perturbe eq ij}) that
\begin{equation}\label{conserved quan}
k_e^2\Phi\propto\left(\mathcal{H}\frac{\delta\phi}{\phi_0'}-\Phi\right)',
\end{equation}
where the prime denotes the derivative with respect to conformal time $\eta\equiv\int\frac{dt}{a}$ and $\mathcal{H}\equiv\frac{a'}{a}$.
During the stage of inflation, we are interested in the modes of quantum fluctuations that are outside the Hubble horizon. Therefore, the terms that are proportional to $k^2$ can be ignored. Then, we find from equation (\ref{conserved quan}) that the comoving curvature perturbation $R_c\equiv\mathcal{H}\frac{\delta\phi}{\phi_0'}-\Phi$ is conserved outside the Hubble horizon. The comoving curvature perturbation $R_c$ is same with that in the standard inflation model. This is due to the fact that $R_c$ does not depends on the wavenumbers $k$, such that the anisotropic effect does not appear explicitly in $R_c$. Thus, we can use the formula (\ref{spectrum delta phi}) to get the primordial power spectrum for $R_c$. It is of the form
\begin{equation}\label{spectrum delta Rc}
\mathcal{P}_{R_c}(\vec{k})=\mathcal{P}_{iso}(k)(1-3b\hat{k}\cdot\hat{n}_z),
\end{equation}
where $\mathcal{P}_{iso}$ is the isotropic power spectrum for $R_c$.

\begin{figure}
\includegraphics[width=8.5 cm]{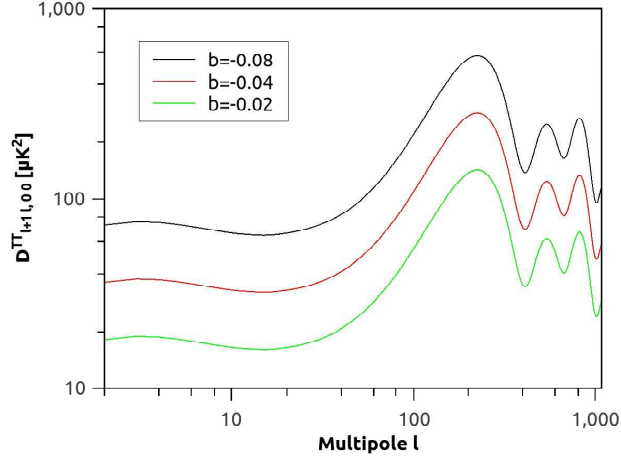}
\caption{The off-diagonal $TT$ correlation coefficients with $m=0$. The black, red and green curve correspond to different Finslerian parameter, i.e. $b=-0.08,-0.04,-0.02$, respectively. The following figures are given by the same convention.}
\label{TT m=0}
\end{figure}
\begin{figure}
\includegraphics[width=8.5 cm]{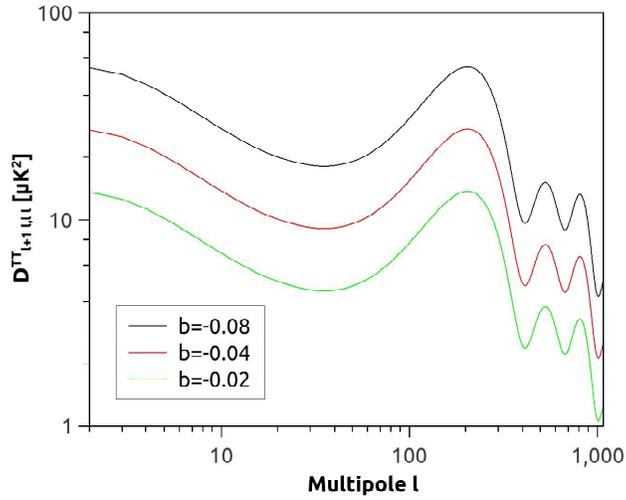}
\caption{The off-diagonal $TT$ correlation coefficients with $m=l$. It demonstrates that the off-diagonal correlation coefficients $C_{ll'}$ depend on $m$.}
\label{TT m=l}
\end{figure}
\begin{figure}
\includegraphics[width=8.5 cm]{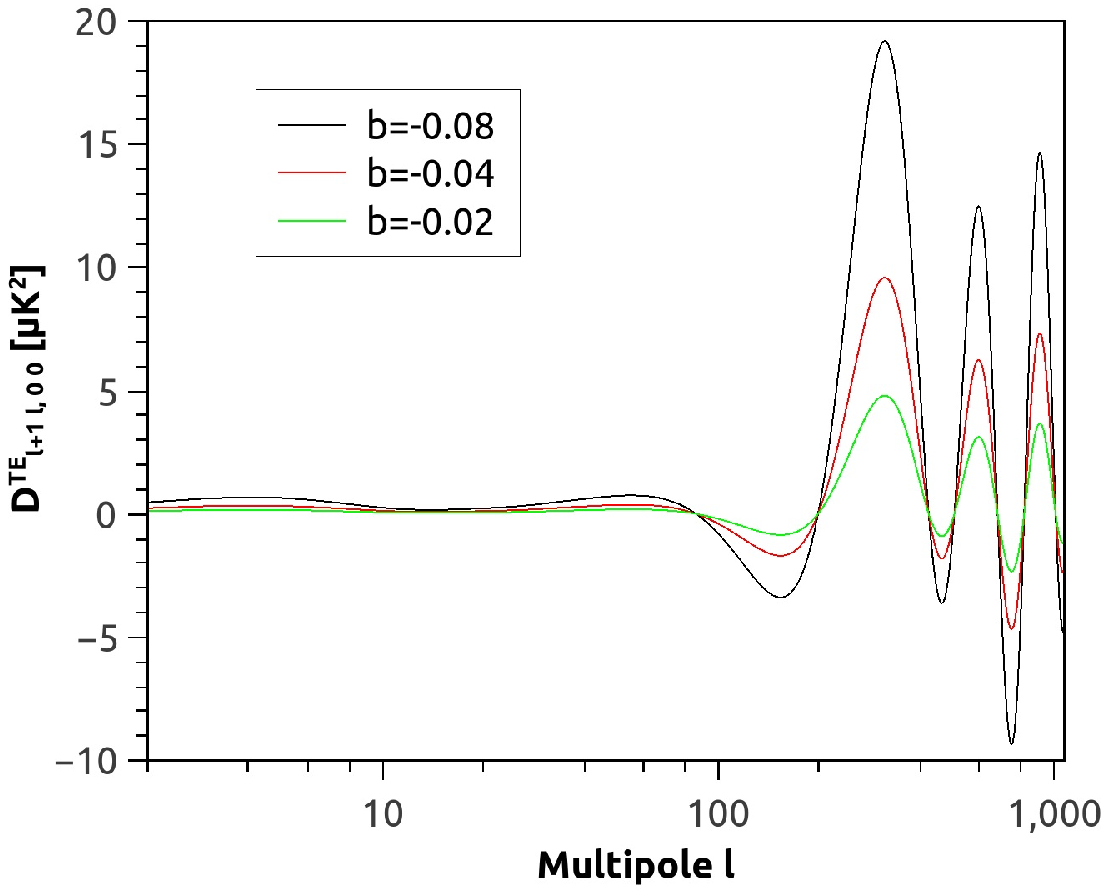}
\caption{The off-diagonal $TE$ correlation coefficients with $m=0$.}
\label{TE m=0}
\end{figure}
\begin{figure}
\includegraphics[width=8.5 cm]{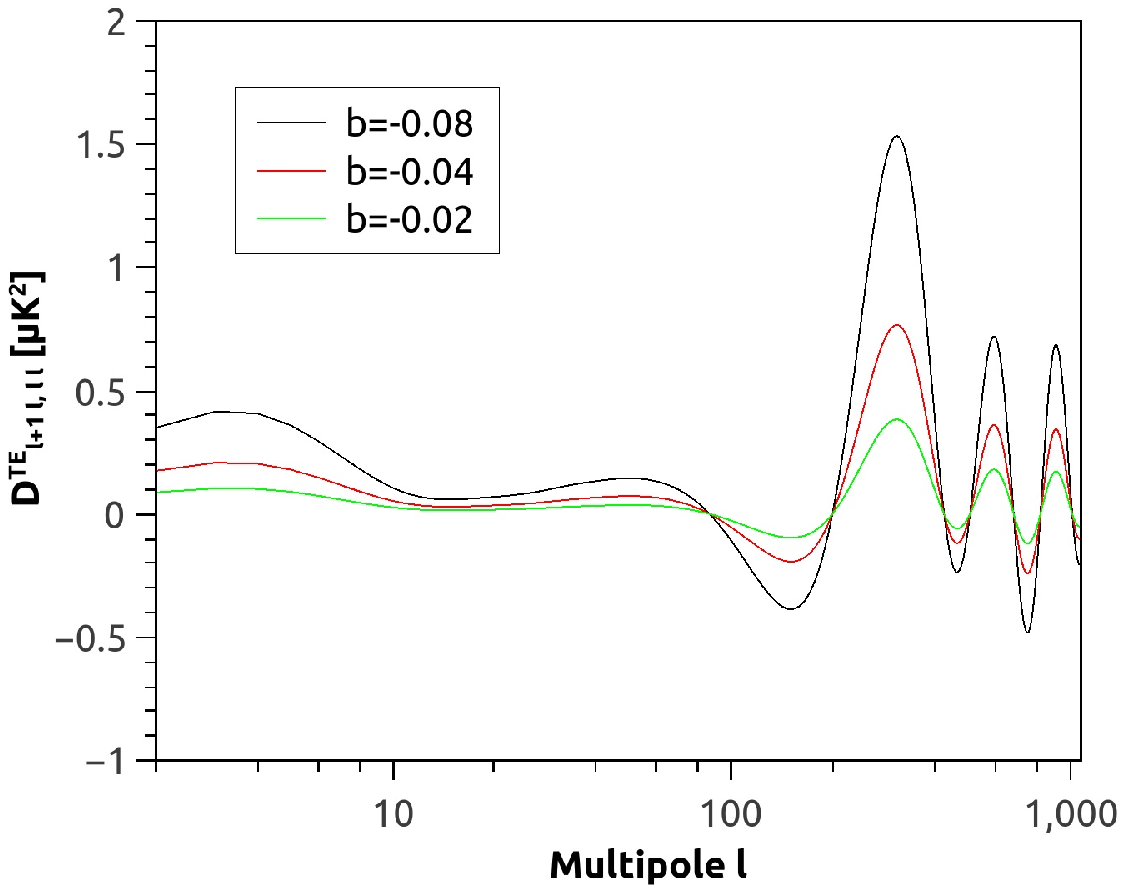}
\caption{The off-diagonal $TE$ correlation coefficients with $m=l$.}
\label{TE m=l}
\end{figure}
\begin{figure}
\includegraphics[width=8.5 cm]{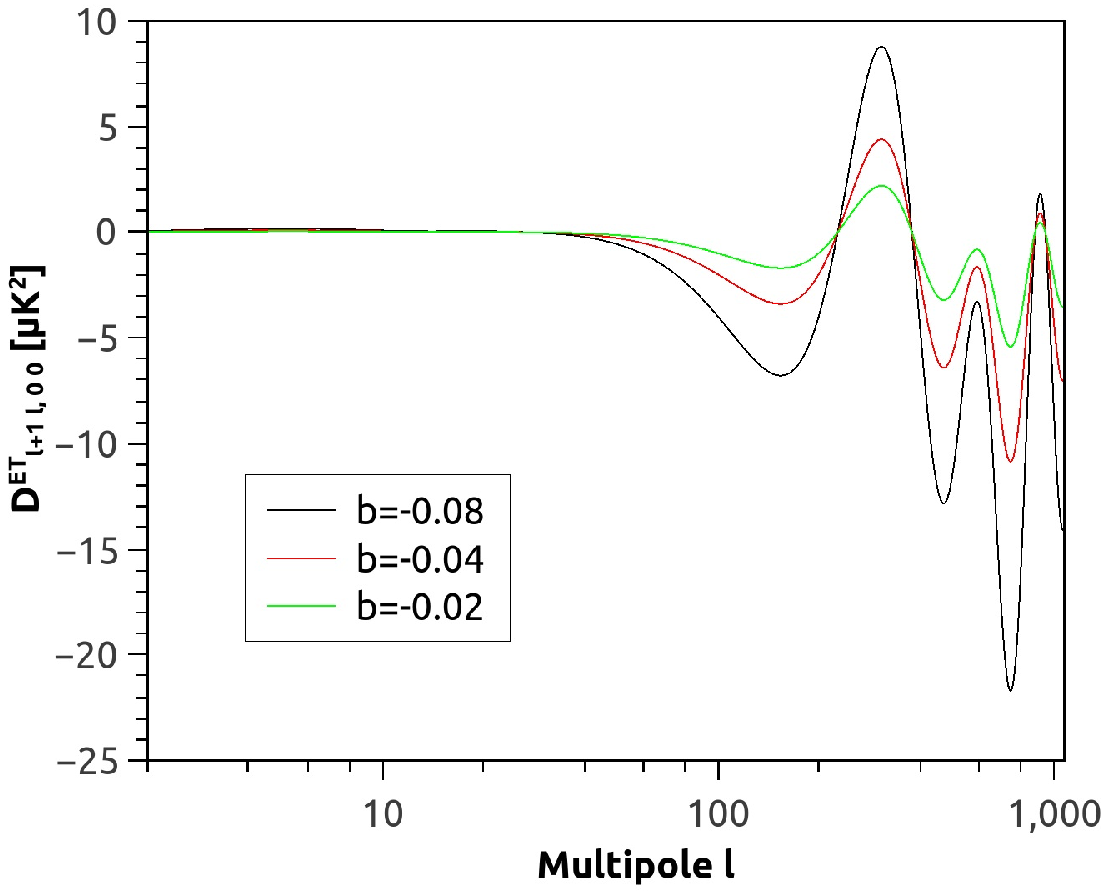}
\caption{The off-diagonal $ET$ correlation coefficients with $m=0$.}
\label{ET m=0}
\end{figure}
\begin{figure}
\includegraphics[width=8.5 cm]{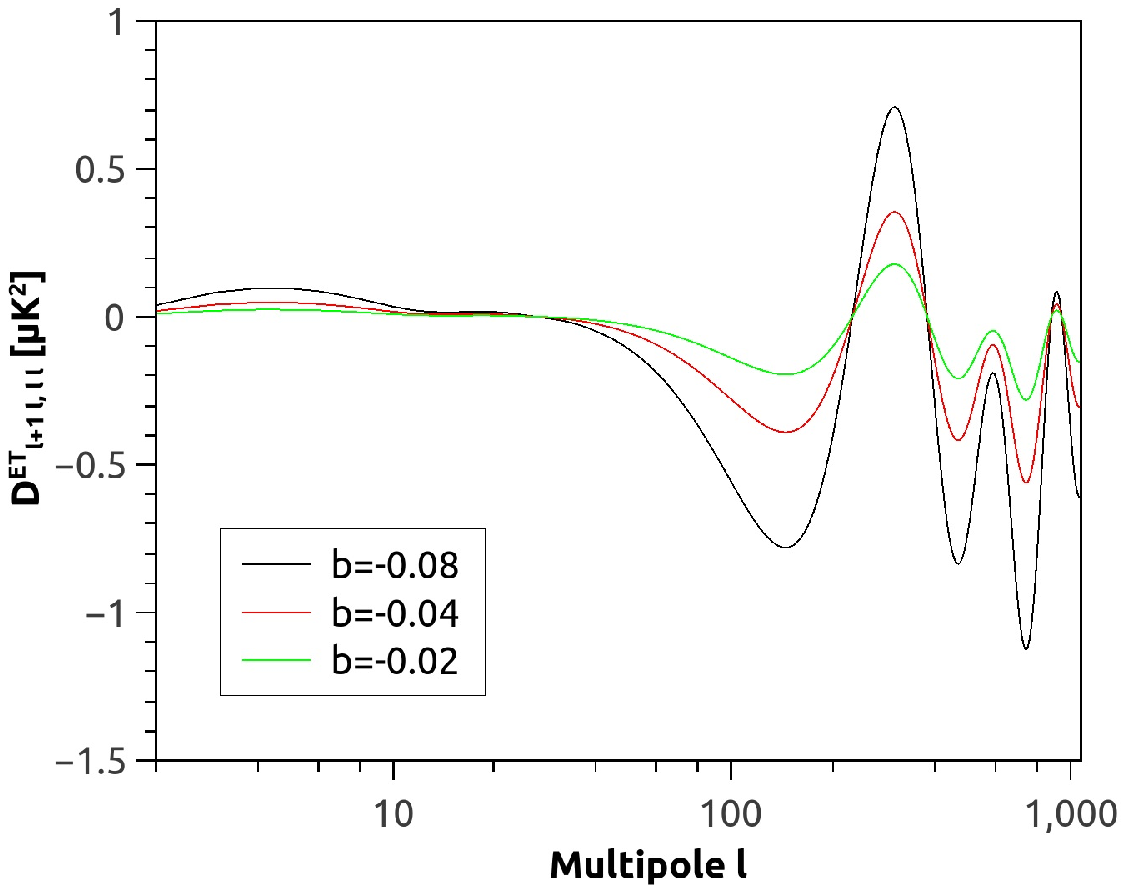}
\caption{The off-diagonal $ET$ correlation coefficients with $m=l$.}
\label{ET m=l}
\end{figure}
\begin{figure}
\includegraphics[width=8.5 cm]{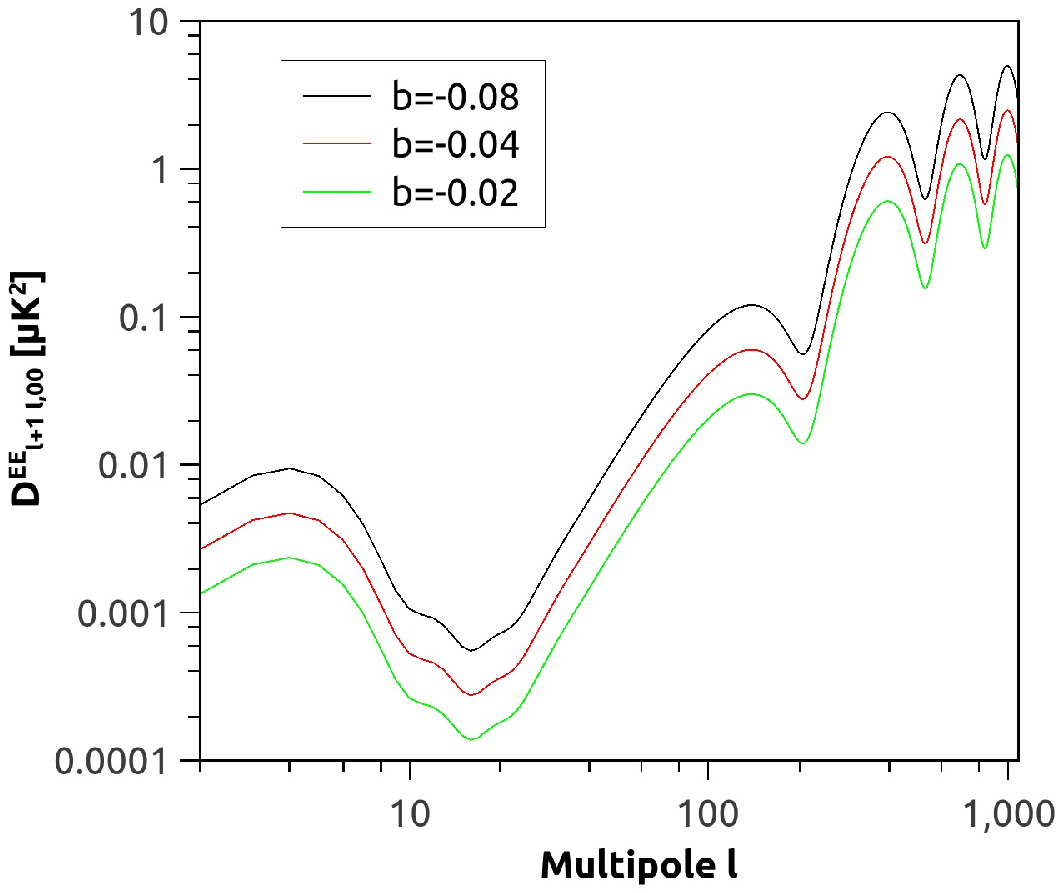}
\caption{The off-diagonal $EE$ correlation coefficients with $m=0$.}
\label{EE m=0}
\end{figure}
\begin{figure}
\includegraphics[width=8.5 cm]{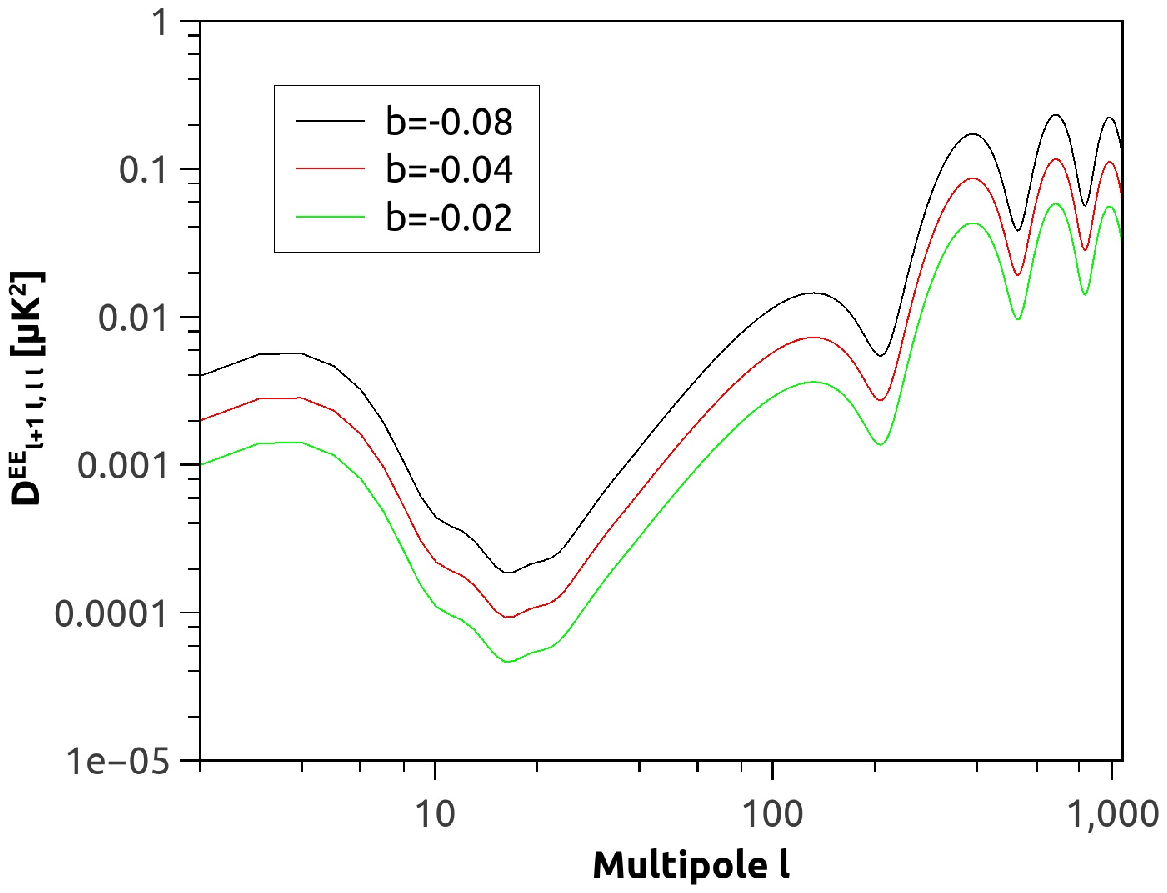}
\caption{The off-diagonal $EE$ correlation coefficients with $m=l$.}
\label{EE m=l}
\end{figure}
\section{Contributions to CMB power spectra from anisotropic inflation model}\label{sec:spectrum}
The anisotropic term in formula (\ref{spectrum delta Rc}) could give off-diagonal angular correlation for the CMB temperature fluctuation and E-mode polarization. The general angular correlation coefficients that describe the anisotropic effect are given by $C_{XX',ll',mm'}$ \cite{off-diagonal}, where $X$ denotes $T$ and $E$, respectively. In our anisotropic model, by making use of the formula (\ref{spectrum delta Rc}), we obtain the CMB correlation coefficients for scalar perturbations as follows
\begin{equation}\label{angular correlation}
C_{XX',ll',mm'}=\int\frac{\ln k}{(2\pi)^3}\Delta_{X,l0}(k)\Delta^\ast_{X',l'0}(k)P_{ll'mm'},
\end{equation}
where
\begin{eqnarray}
P_{ll'mm'}&=&\int d\Omega\mathcal{P}_{R_c}Y^\ast_{lm}Y_{l'm'}\nonumber\\
&=&\mathcal{P}_{iso}\delta_{mm'}\left(\delta_{ll'}-3b\sqrt{\frac{2l'+1}{2l+1}}\mathcal{C}^{l'm}_{10lm}\mathcal{C}^{l'0}_{10l0}\right),
\end{eqnarray}
$\Delta_{X,ls}(k)$ denote the transfer functions and $\mathcal{C}^{lm}_{LMl'm'}$ are the Clebsch-Gordan coefficients. Then, we find from formula (\ref{angular correlation}) that the CMB correlation coefficients $C_{ll'}$ contain diagonal part which means $l'=l$ and off-diagonal part which means $l'=l+1$. Here, we have defined $C_{ll'}\equiv C_{XX',ll',mm'}$. The diagonal part of $C_{ll'}$ that gives the contribution of statistical isotropy equals to the CMB correlation coefficients in standard cosmological model. And the off-diagonal part of $C_{ll'}$ that comes from the anisotropic term $3b\hat{k}\cdot\hat{n}_z$ describes the deviation from statistical isotropy. One should notice that the anisotropic term $3b\hat{k}\cdot\hat{n}_z$ in our model is not invariant under parity flip, $k\rightarrow-k$. Thus, the anisotropic effect in our model does not contribute to the diagonal part of $C_{ll'}$.

Here, by making use of the formula of angular correlation coefficients (\ref{angular correlation}) and the Planck 2013 data \cite{CMB2}, we plot numerical results for the off-diagonal part of $C_{ll'}$. The off-diagonal part of $C_{ll'}$ have two properties that differ from the diagonal part. One is that the $TE$ and $ET$ correlation coefficients are different. The other shows that the off-diagonal part of $C_{ll'}$ depends on $m$. These properties are obvious in the following figures. The off-diagonal $TT,TE,ET,EE$ correlation coefficients for $m=0$ are shown in Fig.\ref{TT m=0}, \ref{TE m=0}, \ref{ET m=0}, \ref{EE m=0}, respectively. And the off-diagonal $TT,TE,ET,EE$ correlation coefficients for $m=l$ are shown in Fig.\ref{TT m=l}, \ref{TE m=l}, \ref{ET m=l}, \ref{EE m=l}, respectively. Here, we have used the mean value of cosmological parameters \cite{CMB2} to give the above figures. And the coefficients $D^{XX'}_{ll',mm'}$ in these figures are defined as $D^{XX'}_{ll',mm'}\equiv \sqrt{l(l+1)l'(l'+1)}C_{XX',ll',mm'}/2\pi$.

The above figures show that the off-diagonal correlation coefficients for $m=0$ have similar shape with the diagonal part of angular correlation coefficients. This is due to the fact that the transfer function $\Delta_l$ approximately equals to $\Delta_{l+1}$ for high $l$. In other words, they are related to nearly same physical scales.

\section{Conclusions and Remarks}\label{sec:conclusion}
In this paper, we have proposed an anisotropic inflation model in Finsler spacetime. The Finsler spacetime (\ref{FRW like}) breaks the rotational symmetry and induces parity violation. The primordial power spectrum was obtained for the quantum fluctuation of inflation field (\ref{spectrum delta phi}). The term $3b\hat{k}\cdot\hat{n}_z$ in the primordial power spectrum $\mathcal{P}_{\delta\phi}$ represents the effect of rotational symmetry breaking. The gravitation field equations of the perturbed Finsler spacetime (\ref{FRW like perturbed}) were presented. We have taken the direction of $y$ to be parallel with the wave vector $\vec{k}$ in momentum space. Thus, the dynamic feature of Finsler spacetime vanishes. The perturbed equations (\ref{perturbe eq tt},\ref{perturbe eq ti},\ref{perturbe eq ii},\ref{perturbe eq ij}) are same with that of the standard inflation model except for replacing wavenumber $k$ with effective wavenumber $k_e$. $k_e$ depends not only on the magnitude of $k$ but also the preferred direction $\hat{n}_z$ that represents rotational symmetry breaking. These field equations guarantee that the comoving curvature perturbation $R_c\equiv\mathcal{H}\frac{\delta\phi}{\phi_0'}-\Phi$ is conserved outside the Hubble horizon. The comoving curvature perturbation $R_c$ is same with the one in the standard inflation model. This is due to the fact that $R_c$ does not explicitly depends on the wavenumber $k$, such that the anisotropic effect does not appear in $R_c$. We have used the primordial power spectrum of $R_c$ (\ref{spectrum delta Rc}) to derive the general angular correlation coefficients $C_{XX',ll',mm'}$ in our anisotropic inflation model. The parity violation feature requires that the anisotropic effect only appears in angular correlation coefficients with $l'=l+1$, and does not contribute to the diagonal $C_{ll}$, i.e. the power spectrum. It means that the off-diagonal part of angular correlation coefficients represents the anisotropic effect. The numerical results for the off-diagonal correlation coefficients show that they depend on $m$, and $TE$ and $ET$ correlation coefficients are different.

\vspace{0.5cm}
\begin{acknowledgments}
We thank Dr. S.-Y. Li and D. Zhao for useful discussions. Project 11375203 and 11305181 supported by NSFC.
\end{acknowledgments}

\end{document}